\documentclass[aps,graphicx,prl,twocolumn,groupedaddress]{revtex4}
\usepackage{amssymb}
\usepackage{epsfig}
\usepackage{graphicx}
\usepackage[centertags]{amsmath}
\usepackage{latexsym}

\newcommand{\ket}[1]{\left|#1\right>}
\newcommand{\bra}[1]{\left<#1\right|}

\begin{document}
\title{Resonant X-rays at the Cu L-edge as a momentum dependent probe of the magnetic excitation spectrum in cuprates}
\author{L.J.P. Ament$^1$, G. Ghiringhelli$^2$, M. Moretti Sala$^2$, L. Braicovich$^2$ and J. van den Brink$^{1,3,4}$}
\affiliation{
$^1$Institute-Lorentz for Theoretical Physics,  Universiteit  Leiden, 2300 RA Leiden,The Netherlands\\
$^2$INFM/CNR Coherentia and Soft -- Dipartimento di Fisica, Politecnico di Milano, Piazza Leonardo da Vinci 32, 20133 Milano, Italy\\
$^3$Institute for Molecules and Materials, Radboud Universiteit Nijmegen, 6500 GL Nijmegen, The Netherlands\\
$^4$Stanford Institute for Materials and Energy Sciences, Stanford University and SLAC National Accelerator Laboratory, Menlo Park, CA 94025.
}
\date{\today}

\begin{abstract}
We show that, contrary to common lore, in resonant inelastic x-ray
scattering (RIXS) at the copper L- and M-edge direct spin-flip
scattering is in principle allowed. We demonstrate how this
possibility can be exploited to probe the momentum dependent
magnetic properties of cuprates such as the high T$_{\rm c}$
superconductors and compute in detail the relevant local and
momentum dependent magnetic scattering amplitudes, which we compare
to the elastic and $dd$-excitation scattering intensities. For
cuprates these results put RIXS as a technique on the same footing
as neutron scattering.
\end{abstract}

\maketitle

{\it Introduction.} In recent years the experimental technique of
resonant inelastic x-ray scattering (RIXS) has made tremendous
progress in terms of energy and momentum
resolution~\cite{Schulke07,Kotani01,Kuiper98,Ghiringhelli04,Veenendaal06,Chiuzbaian05,Ghiringhelli09,Hasan00,Kim02,Markiewicz06,Hill08}. RIXS is particularly apt to probe the properties of strongly correlated electrons, for instance
the ones of the transition metal oxides~\cite{Schulke07,Kotani01}.
With an incoming x-ray of energy $\omega_{in}$ first an electron is
resonantly excited from a core level into the valence shell.
Subsequently one measures the energy $\omega_{out}$ of the outgoing
x-ray resulting from the recombination of the core hole with a
valence electron. Depending on the resonance that the experiment is
performed at, $\omega_{in}$ corresponds to the transition metal
K-edge ($1s \rightarrow 4p$),  L-edge ($2p \rightarrow 3d$) or
M-edge ($3p \rightarrow 3d$). Compared to the many other
photon scattering techniques, RIXS has the advantage that there is
no core hole present in the final state, so that the energy lost by
the scattered photon at a transition metal L- or M-edge is
directly related to electronic excitations within the strongly
correlated $3d$ valence bands. The chemical selectivity and bulk
sensitivity of RIXS allows the study of the electronic and magnetic
properties of, for example, complex and nanostructured materials
that might be inaccessible with non-resonant techniques.

RIXS has been much used in transition metal oxides to study local
transitions, such as $dd$-excitations in
cuprates~\cite{Kuiper98,Ghiringhelli04,Veenendaal06} and spin
flips~\cite{Chiuzbaian05,Ghiringhelli09} in NiO. Although
interesting in themselves, these experiments do not exploit a unique
feature of RIXS: it allows the measurement of the {\it dispersion}
of excitations by determining both momentum change and energy loss
of the scattered x-ray photons. Such a capability is far beyond the
possibilities of traditional low energy optical techniques, which
are constrained to zero momentum transfer because, as opposed to
high energy x-rays, photons in the visible range carry negligible
momentum. This asset has already been exploited to determine the
momentum dependence of charge~\cite{Hasan00,Kim02,Markiewicz06},
bimagnon~\cite{Hill08,Brink07,Nagao07,Forte08b} and orbital 
excitations~\cite{Forte08a,Ulrich} at Cu K and L$_3$ edges. In this Letter we 
show that, contrary to common lore~\cite{Kuiper98,DeGroot98,Kotani01,Veenendaal06}, RIXS
at the copper L-edge is also a very powerful probe of single-magnon
excitations and their dispersions. This is an important result
because it puts the technique of L-edge RIXS on, for instance, the
high T$_c$ superconductors, on the same footing as neutron
scattering --with the considerable advantage that x-ray scattering
requires only small sample volumes. 

In the following we will first determine the local spin-flip cross
section for a copper $d^9$ ion in a tetragonal crystal field. This
is the familiar case encountered in numerous cuprates, where the
only unquenched magnetic moment in the system is the one of a hole
occupying a $x^2$-$y^2$ orbital. The important observation is that
it strongly depends on the spatial orientation of the copper spin
whether this local spin-flip process is forbidden or not. In the
second part of the Letter we determine the momentum ({\bf q}-)
dependence of the magnon cross section for a spin system with
collective response. As an example we consider the Heisenberg
antiferromagnet, where we find a vanishing of the magnon scattering
intensity around the center of the Brioullin zone proportional to
$|{\bf q}|$ and strongly peaking of it around the antiferromagnetic
wave vector.

{\it Local spin-flip scattering at Cu L-edge.} From the viewpoint of
inelastic magnetic scattering it appears that  RIXS and neutron
scattering are very different techniques. It is easy to show, for
instance, that in transition metal K-edge RIXS single spin-flip
scattering is forbidden~\cite{Brink07} because of the absence of
spin-orbit coupling in the intermediate state. Ever since the
seminal work of Kuiper and coworkers~\cite{Kuiper98}, more than a
decade ago, it is believed that also at the copper L- and M-edge
spin-flip scattering is not allowed for Cu$^{2+}$ in a tetragonal
crystal field, unless the spin-flip excitation is accompanied by a
$dd$-excitation~\cite{DeGroot98,Kotani01,Veenendaal06}.  Based on a
symmetry analysis of the wavefunction of the copper hole,
Ref.~\onlinecite{DeGroot98} states that ``the reason is that the
$x^2$-$y^2$ state, a linear combination of atomic $Y_{2,2}$ and
$Y_{2,-2}$ states, does not allow a direct spin-flip transition''.
The observation that the spin-flip excitations are intrinsically
entangled with $dd$-excitations implies that mapping out momentum
dependencies of magnetic excitations with L-edge RIXS would be a
hopeless endeavor. As will be clarified shortly, the
$dd$-excitations act as a momentum sink, which would limit the
information that can be gained from RIXS in this case to {\it
momentum averaged properties} of the magnetic excitations,
preempting the possibility to observe single magnon dispersions.

We will show in the following, however, that the symmetry analysis
on which these assertions rely~\cite{DeGroot98,Veenendaal06} is
incomplete because restricted to directions of the spin moment along
an axis that is orthogonal to the $x^2$-$y^2$ orbital. In fact we
show that {\it for any other spin orientation direct spin-flip
scattering is allowed}. This includes in particular the N\'eel
ordered high T$_c$ cuprates, where the magnetic moment lies in the
plane of the $x^2$-$y^2$ orbital: for example in La$_2$CuO$_4$,
Sr$_2$CuO$_2$Cl$_2$ and (CaSr)CuO$_2$,\cite{LCOspin, CSCOspin} spins
point along the [$x,y,z$]=[110] direction and in Nd$_2$CuO$_4$ along
[100] and [010] in alternating planes.\cite{NCOspin}

The dependence of the direct spin-flip scattering amplitude on
photon polarization, scattering angle and momentum transfer can be
computed from the Kramers-Heisenberg expression~\cite{Blume85}:
$    A_{fi} \propto \sum_n \bra{f} \hat{D} \ket{n} \bra{n} \hat{D} \ket{i} /({\omega_{\rm det}-E_n +i \Gamma})$,
where $A_{fi}$ is the scattering amplitude from initial state $i$ to
final state $f$,  $\omega_{\rm res}$ is the resonant energy (around
930 eV at the copper L$_3$-edge), $\hat{D}$ the polarization
dependent dipole operator,  $\omega_{det} \equiv \omega_{in} -
\omega_{res}$ is the detuning away from the resonance, $\ket{n}$ the
intermediate state with energy $E_n$ (measured with respect to the
resonance energy) and $\Gamma$ the core-hole lifetime. At the copper
L-edge we are dealing with the local electronic process 2$p^6$3$d^9
\rightarrow$ 2$p^5$3$d^{10} \rightarrow$ 2$p^6$3$d^{9*}$, where $*$
denotes an excited state with a $dd$-excitation and/or spin flip. At
the L$_3$ resonance the intermediate states $\ket{n}$ are just the
multiplets corresponding to the four $J^z=L^z+S^z$ states of the
spin-orbit coupled 2$p_{3/2}$ core-hole so that the calculation
becomes relatively simple because all paths from the ground state to
a given final state interfere with equal weight.

\begin{figure}
\begin{center}
\includegraphics[width=.85\columnwidth]{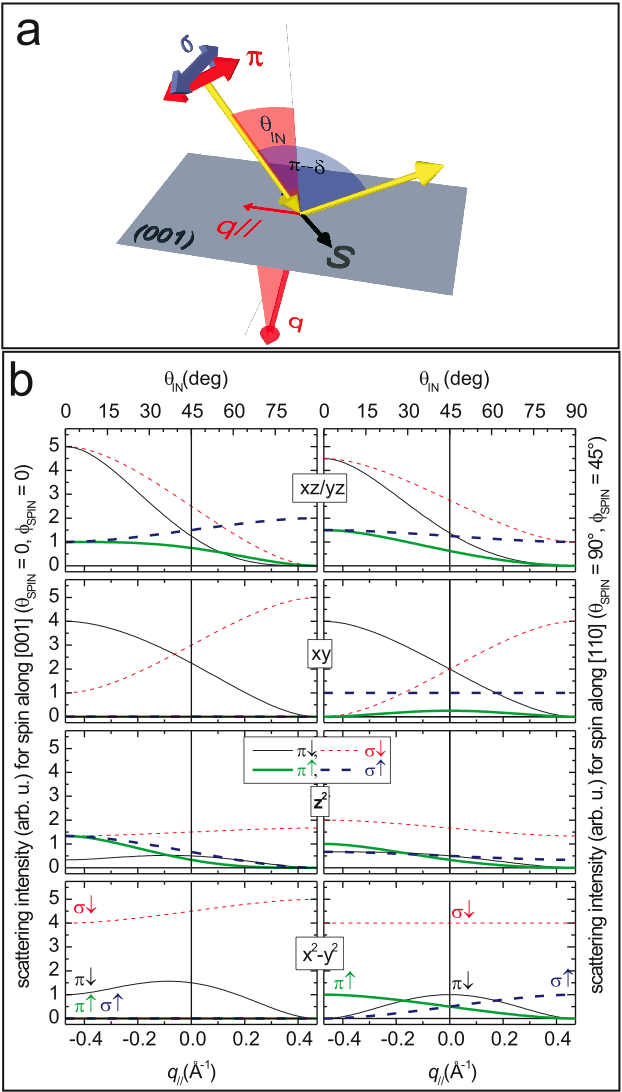}
\caption{ The Cu$^{2+}$ L$_3$ local RIXS cross section for $\sigma$
and $\pi$ polarization of the incident beam calculated in the single
ion model. The experimental geometry is sketched in panel a: the
scattering plane is (100), the scattering angle is 90$^o$, the
incident photons impinge at an angle $\theta_{IN}$ to the [001]
crystallographic direction ($c$ axis). The spin orientation is
varied in the calculation. In panel b, all the possible
$dd$-excitations plus the spin-flip and the elastic scattering final
states are considered, vs $\theta_{IN}$ or alternatively the
in-plane transferred momentum $q_{\parallel}$. We notice that along
the [100] direction the nuclear Brillouin zone boundary is in
$q_{\parallel}\approx $ 0.826$\hbar$ \AA$^{-1}$ for $a$ = 3.8 \AA\;
in-plane parameter, typical of cuprates: at 90$^{\circ}$ scattering
Cu L$_3$ RIXS can explore half of the reciprocal space, but going to
backscattering geometry $q_{\parallel}$ grows considerably and
almost all the Brillouin zone can be covered.  When the spin is
along [001] the spin flip cross section vanishes (bottom left), but
not when the spin is along [110] (bottom right). } \label{fig:1}
\end{center}
\end{figure}

It is easy to see that direct spin-flip excitations are forbidden if
the spin of the hole in the
$x^2$-$y^2$=$(Y_{2,2}+Y_{2,-2})/\sqrt{2}$ initial state is aligned
along [001], which is the situation considered
previously~\cite{DeGroot98,Veenendaal06}. In the first step of the
RIXS process a dipole allowed $2p \rightarrow 3d$ transition creates
a core hole in a linear combination of $Y_{1,1}$ and $Y_{1,-1}$,
while conserving the spin. In this intermediate state the spin-orbit
coupling of the core-hole ${\bf L} \cdot {\bf S} = L^z S^z + (L^+S^-
+ L^-S^+)/2$ can cause a spin-flip $S^-$ (or $S^+$) in combination
with a raising (or lowering)  $L^+$ (or $L^-$) of the orbital
moment.  In either case a $Y_{1,0}$ core-hole state with reversed
spin is the result~\cite{DeGroot98,Veenendaal06,Veenendaal_private}.
The last step to end up in a final state with only a spin-flip
excitation, requires the optical decay of the $Y_{1,0}$ $2p$
core-hole into a $(Y_{2,2}+Y_{2,-2})/\sqrt{2}$ $3d$ valence band
hole. But this transition is dipole forbidden because it requires
$\Delta L^z =2$, which thus forbids direct spin-flip scattering
completely. According to the same argument, a local spin-flip is
allowed if the hole in the final state is any another orbital than
$x^2$-$y^2$.

\begin{figure}
\begin{center}
\includegraphics[width=\columnwidth]{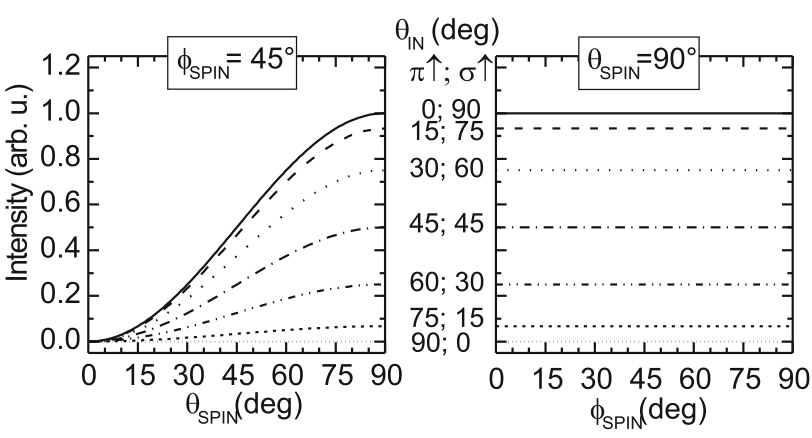}
\end{center}
\caption{The dependence of scattering cross section for elastic
($x^2$-$y^2$)$\downarrow$ and spin-flip ($x^2$-$y^2$)$\uparrow$
final states on the atomic spin orientation for selected cases of
scattering geometry given by varying $\theta_{\textrm{IN}}$ and
fixed $\phi_{\textrm{IN}}=0$.} \label{fig:2}
\end{figure}

The situation changes drastically when the local magnetic moment is
oriented in the $x$-$y$ plane: we will show that now direct
spin-flip excitations become allowed. This is best illustrated by a
direct calculation of the RIXS amplitudes in the different channels
for  Cu$^{2+}$ in a tetragonal crystal field. We did so for the
scattering geometry sketched in Fig.~\ref{fig:1}a, with  fixed
scattering angle of 90$^o$ and $\pi$ ($\sigma$) linear polarization
of the incident photons parallel (perpendicular) to the scattering
plane. In this geometry $\theta_{\textrm{IN}}$ is the azimuthal
angle between incident beam and [001] axis.  In Fig.~\ref{fig:1}b
the exact polarization and momentum dependent RIXS matrix elements
for all possible final states and for an initial magnetic moment
either along [001] (left panels) or along [110] (right panels) are
shown. The computed spin flip and $dd$-excitation cross section
start from a local ground state (x$^2$-y$^2$)$\downarrow$, so that
(x$^2$-y$^2$)$\uparrow$ is a final state with a spin flip excitation
only.  On the bottom axis of Fig.~\ref{fig:1}b is the momentum
transfer in the experimental geometry at $\omega_{in}$ = 930 eV.
From the lower panels at the left and right it is clear that for a
spin along [110] the spin flip cross section is allowed for both
$\sigma$ and $\pi$ polarizations, whereas it is in all cases
forbidden for a spin along [001]. It is interesting to note that for
the $\sigma$ polarization the elastic peak is more than 4 times
stronger than the spin-flip scattering channel, whereas for $\pi$
the two intensities are similar. The direct spin-flip and
elastic cross section for a generic spin direction, characterized by
the Euler angles ($\theta_{\textrm{spin}},\phi_{\textrm{spin}}$) are
shown in Fig.~\ref{fig:2} for a number of azimuthal angles
$\theta_{\textrm{IN}}$.

The upshot of the numerical  results above can easily be understood on the basis of a symmetry argument. If the spin of the $x^2$-$y^2$ hole points along the $x$-axis, it is in the spin state $(\ket{\uparrow} + \ket{\downarrow})/\sqrt{2}$, corresponding to $ S^x =1/2$. In the intermediate $2p_{3/2}$ core-hole state the diagonal part of the spin-orbit coupling, $L^z S^z$,
causes a transition of this spin state into $(\ket{\uparrow} -
\ket{\downarrow})/\sqrt{2}$ (corresponding to $S^x=-1/2$), while the
angular part of the core-hole wavefunction stays in a linear
combination of $Y_{1,1}$ and $Y_{1,-1}$. The transition of the
core-hole back into the $3d$ $x^2$-$y^2$ orbital is therefore dipole
allowed while at the same time the spin along the $x$-axis is
flipped.

{\it Momentum dependence of magnon cross section.} We now wish to
generalize the cross section from local spin-flips to collective
magnetic excitations, which are characterized by their momentum
quantum number $\bf q$. There are several ways to compute the $\bf
q$ dependence of this cross section, but a particularly transparent
one is by expanding the Kramers-Heisenberg scattering amplitude
formally in a power series of the intermediate state Hamiltonian
$H_{\rm int}$:
$    A_{fi} = \sum_{l=0}^{\infty}\bra{f} \hat{D} (H_{\rm int})^l \hat{D} \ket{i}/  \Delta^{l+1} $,
with $\Delta = \omega_{\rm det}+i\Gamma$, which is the starting
point for the ultra-short core-hole life-time expansion for
RIXS\cite{Brink06,Ament07,Forte08b}. The dipole operator $\hat{D}=\hat{D}_0+\hat{D}_0^\dagger$, now allows for a direct creation of a spin-flip upon de-excition The corresponding amplitude $r_{\bf q}$ depends on polarization, transferred momentum, scattering geometry and initial orientation of the magnetic moment as clarified above and shown in Figs.~\ref{fig:1} and~\ref{fig:2}.
We thus obtain
\begin{eqnarray}
    \hat{D}_0  = \sum_{i} e^{i {\bf q}_{\rm out}\cdot {\bf R}_i} \left(1-r_{\bf q}+r_{\bf q} S^x_i \right)  d^{\dag}_{i} p^{\phantom{\dag}}_{i}
    + e^{-i {\bf q}_{\rm in}\cdot {\bf R}_i}  p^{\dag}_{i} d^{\phantom{\dag}}_{i} \nonumber
\end{eqnarray}
where $d^{\dag}_{i}$ and $p^{\dag}_{i}$ create a $3d$ valence hole and $2p$ core-hole, respectively, and we have taken the direction of the initial magnetic moment as the spin quantization $z$-axis.   By doing so all magnetic effects of the intermediate state core-hole Hamiltonian have been taken into account so that they can be integrated out and the scattering amplitude is
$A_{fi} = \Delta^{-1} \bra{f} \sum_i e^{i {\bf q}\cdot {\bf R}_i} \left(1-r_{\bf q}+r_{\bf q} S^x_i \right)  \ket{i}$
with ${\bf q} = {\bf q}_{\rm out} - {\bf q}_{\rm in}$. The inelastic part of the magnetic scattering amplitude finally reduces to
$A_{fi} = \frac{ r_{\bf q}}{\Delta} \bra{f}  S^x_{\bf q}  \ket{i}$.

\begin{figure}
\begin{center}
\includegraphics[width=.8\columnwidth]{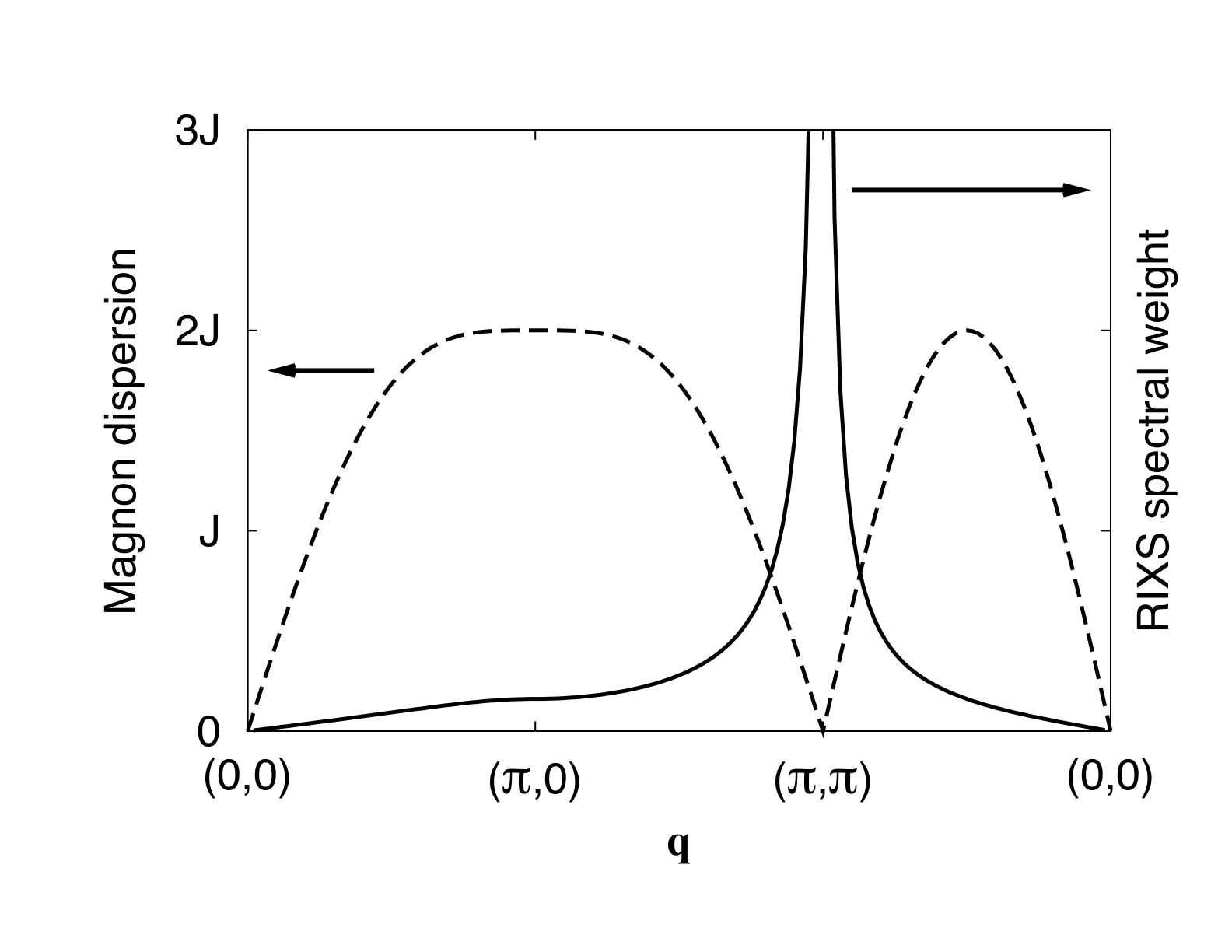}
\caption{Momentum dependence of the single magnon spectral weight
for RIXS at the copper L$_3$ edge (solid line) and the magnon
dispersion (dashed line) for a simple 2D Heisenberg
Hamiltonian\label{fig:singlemagnon}} \vspace{-0.3cm}
\end{center}
\end{figure}

It is instructive to compute with this generic expression the single
magnon RIXS spectrum for, for instance, the antiferromagnetic
two-dimensional Heisenberg model, given by the Hamiltonian $H= J
\sum_{\langle ij \rangle} {\bf S}_i \cdot {\bf S}_j$. Introducing
Holstein-Primakoff bosons for $i$ in sublattice $A$, $S^z_i \mapsto
1/2-a^{\dag}_i a^{\phantom{\dag}}_i$, $S^+_i \mapsto
a^{\phantom{\dag}}_i$ and $S^-_i \mapsto a^{\dag}_i$. For $j \in B$,
we have $S^z_j \mapsto b^{\dag}_j b^{\phantom{\dag}}_j-1/2$, $S^+_j
\mapsto b^{\dag}_j$ and $S^-_j \mapsto b^{\phantom{\dag}}_j$.
Adopting linear spin wave theory one finds after a Fourier and a
Bogoliubov transformation the magnon scattering amplitude
$    A_{fi} = \sqrt{N/2}  (u_{\bf q}-v_{\bf q}) \bra{f} \alpha^{\phantom{\dag}}_{\bf q} + \beta^{\phantom{\dag}}_{\bf q} + \alpha^{\dag}_{-{\bf q}} + \beta^{\dag}_{-{\bf q}} \ket{i} /{\Delta }$
with $N$ the total number of sites and the Bogoliubov transformed
boson operators $\alpha^{\phantom{\dag}}_{\bf k} =
u^{\phantom{\dag}}_{\bf k} a^{\phantom{\dag}}_{\bf k} +
v^{\phantom{\dag}}_{\bf k} b^{\dag}_{-{\bf k}}$ and
$\beta^{\phantom{\dag}}_{\bf k} = u^{\phantom{\dag}}_{\bf k}
b^{\phantom{\dag}}_{\bf k} + v^{\phantom{\dag}}_{\bf k}
a^{\dag}_{-{\bf k}},$ where $u^2_{\bf k} - v^2_{\bf k} = 1$, $u_{\bf
k} = \sqrt{ \frac{J}{\omega_{\bf k}}+\frac{1}{2}}$, $\omega_{\bf
k}=2J{\sqrt{1 - \gamma_{\bf k}^2}}$ and $\gamma_{\bf k} =(\cos k_x + \cos k_y)/2$. The resulting zero-temperature
single magnon spectrum is shown in Fig.~\ref{fig:singlemagnon}. At
${\bf q} = (0,0)$ the magnon scattering amplitude vanishes because
in this situation the scattering operator is proportional to the
total spin in the $x$-direction $S^x_{\rm tot}$, which does not
cause inelastic processes because this operator commutes with the
Heisenberg Hamiltonian. For small transferred momenta, $|{\bf q}|
\rightarrow  0$, the magnon scattering intensity vanishes as $
\omega_{ {\bf q}} / 4J$. We also observe that the magnon cross
section diverges at ${\bf q} = (\pi,\pi)$ as $4J/\omega_{\bf q}$,
similar to the neutron scattering form factor. This divergence is
due to the RIXS photons scattering on spin fluctuations: at ${\bf q}
= (\pi,\pi)$ the scattering operator is proportional to the
staggered spin along the $x$-axis $S^x_{\rm stag}$, so that the
total, energy integrated, scattering intensity $\int d \omega \sum_f
|A_{fi}|^2 \delta (\omega-\omega_{fi}) \propto \langle (S^x_{\rm
stag})^2\rangle$ and the inelastic scattering intensity is
proportional to the variance $\langle (S^x_{\rm stag})^2\rangle -
\langle S^x_{\rm stag}\rangle^2$. In the 2D Heisenberg model the
expectation value of these spin fluctuations is proportional to
$N^{3/2}$.  

Using the same formalism, we can compute the $\bf q$ dependent
scattering amplitude of a spin-flip entangled with a
$dd$-excitation. If the local spin-flip operator is $S^-_i$ and the
operator corresponding to the $dd$-transiton is $T^+_i$, the
inelastic scattering amplitude is $A_{fi} \propto \bra{f} \sum_i
e^{i {\bf q}\cdot {\bf R}_i}  S^-_i T^+_i  \ket{i} =  \bra{f}
\sum_{\bf k} S^-_{{\bf k}} T^+_{{\bf k}-{\bf q}}  \ket{i}$. Clearly
part of the momentum is absorbed by the $dd$-excitation, so that
RIXS measures a momentum convolution of the two excitations. In
particular, the magnetic scattering amplitude looses all $\bf q$
dependence if the $dd$-excitation is dispersionless, exemplifying
that in order to determine magnon dispersions the presence of a {\it
direct} spin-flip process is essential.

{\it Conclusions.} Depending on the spatial orientation of the
copper spin, local spin-flip process for RIXS at the L$_3$ edge can
be forbidden or allowed. This makes RIXS a very sensitive probe of
the orientation of the local magnetic moment. In typical cuprates
direct spin-flip scattering is allowed and for this case we
determined the spin-flip and magnon cross section, which turns out
to be strongly momentum and polarization dependent. Our theory holds
at both the copper L- and M-edges. At the M-edge ($\omega_{\rm in} \approx 75$eV) the photon momentum is small, so that only magnons in a very small portion of the Brioullin zone can be probed.  But at the copper L$_3$ edge the X-ray photon carries a momentum $|{\bf q}_{\rm in}| \sim 0.47 \hbar$ \AA$^{-1} $, which is in a typical cuprate large enough to observe magnetic excitations in almost all of the Brillouin zone. Indeed in recent  high resolution RIXS experiments on La$_2$CuO$_4$ single magnon scattering features can be discerned~\cite{Braicovich:unp2}. 
Thus, at least for high T$_c$ superconductors, L-edge RIXS can be placed on the same footing as neutron scattering --with the additional great advantage that for photon scattering only small sample volumes are required so that the measurement of the spin dynamics of thin films,
oxide-heterostructures and other nanostructures comes now within experimental reach.

{\it Acknowledgments.} We thank Michel van Veenendaal, Tom
Devereaux, Maurits Haverkort, Marco Grioni and George Sawatzky for
stimulating discussions. This work is supported by the Department of
Energy, Office of Basic Energy Sciences under contract
DE-AC02-76SF00515 and the Dutch Science Foundation FOM.


\begin{references}
\bibitem{Schulke07}W. Sch\"ulke, Electron Dynamics by Inelastic X-Ray Scattering, Oxford University Press, Oxford, (2007).
\bibitem{Kotani01} For a review see A. Kotani and S. Shin, Rev. Mod. Phys. {\bf 73}, 203 (2001).
\bibitem{Kuiper98} P. Kuiper {\it et al.}, Phys. Rev. Lett. {\bf 80}, 5204 (1998).
\bibitem{Ghiringhelli04}G. Ghiringhelli {\it et al.}, Phys. Rev. Lett. {\bf 92}, 117406 (2004).
\bibitem{Veenendaal06} M. A. van Veenendaal, Phys. Rev. Lett. {\bf 96}, 117404 (2006).
\bibitem{Chiuzbaian05} S. G. Chiuzb\u aian {\it et al.}, Phys. Rev. Lett. {\bf 95}, 197402 (2005).
\bibitem{Ghiringhelli09} G. Ghiringhelli {\it et al.},  Phys. Rev. Lett. {\bf 102}, 027401 (2009).
\bibitem{Hasan00} M. Z. Hasan {\it et al.}, Science {\bf 288}, 1811 (2000).
\bibitem{Kim02} Y.J. Kim {\it et al.}, Phys. Rev. Lett. {\bf 89}, 177003 (2002).
\bibitem{Markiewicz06} R.S. Markiewicz and A. Bansil, Phys. Rev. Lett. {\bf 96}, 107005 (2006). 
\bibitem{Hill08} J.P. Hill {\it et al.}, Phys. Rev. Lett. {\bf 100}, 097001 (2008).
\bibitem{Brink07} J. van den Brink, Europhys. Lett. {\bf 80}, 47003 (2007).
\bibitem{Nagao07} T. Nagao and J.I. Igarashi, Phys. Rev. B {\bf 75}, 214414 (2007).
\bibitem{Forte08b} F. Forte, L.J.P. Ament and J. van den Brink, Phys. Rev. B. {\bf 77}, 134428 (2008).
\bibitem{Forte08a} F. Forte, L. Ament and J. van den Brink, Phys. Rev. Lett. {\bf 101}, 106406 (2008).
\bibitem{Ulrich} C. Ulrich, private communication.
\bibitem{DeGroot98} F.M.F de Groot, P. Kuiper and G.A. Sawatzky, Phys. Rev. B {\bf 57}, 14584 (1998).
\bibitem{LCOspin} D. Vaknin {\it et al.}, Phys. Rev. Lett. {\bf 58}, 2802 (1987).
\bibitem{CSCOspin} D. Vaknin {\it et al.}, Phys. Rev. B {\bf 39}, 9122 (1989).
\bibitem{NCOspin} S. Skanthakumar {\it et al.}, Physica C {\bf 160}, 124 (1989).
\bibitem{Blume85} M. Blume, J. Appl. Phys. {\bf 57}, 3615 (1985).
\bibitem{Veenendaal_private} M. van Veenendaal, private communication.
\bibitem{Brink06} J. van den Brink and M. van Veenendaal, Europhysics Letters {\bf 73}, 121 (2006).
\bibitem{Ament07}L.J.P. Ament, F. Forte and J. van den Brink, Phys. Rev. B {\bf 75}, 115118 (2007).
\bibitem{Braicovich:unp2} L. Braicovich {\it et al.} unpublished.
\end{references}
\end{document}